\documentstyle[12pt]{article}
\newcommand{\beq}{\begin{equation}}
\newcommand{\eeq}{\end{equation}}

\newcommand{\beqn}{\begin{eqnarray}}
\newcommand{\eeqn}{\end{eqnarray}}

\newcommand{\ba}{\bar \alpha}
\newcommand{\bb}{\bar \beta}

\begin{document}
\begin{center}
{\bf Comment on ``Proton Polarization Shifts\\
 in Electronic and Muonic Hydrogen''\\}
\vspace*{1cm}
I.B. Khriplovich and R.A. Sen'kov \\
{\it Budker Institute of Nuclear Physics,\\}
{\it 630090 Novosibirsk, Russia,\\}
{\it and Novosibirsk University }
\end{center}

\vspace*{.5cm}
\begin{abstract}
{\small We demonstrate that criticisms  concerning our work nucl-th/9704043,
made in the recent preprint hep-ph/9903352, are unfounded. The improvements
over our results claimed in hep-ph/9903352 are in fact spurious, being
based mainly on misunderstandings and on dubious theoretical models.}
\end{abstract}

\vspace*{1cm}
In~\cite{ks} (see also~\cite{ks1}) we found the correction to the 
hydrogen Lamb shift due to the proton electric and magnetic 
polarizabilities. This our result was 
cricized recently in~\cite{ro}. One cannot agree with these
criticisms.

To make our objections more comprehensible, let us   
present at first the key points of our derivation~\cite{ks}. We started
it with the electron-proton forward scattering amplitude:
\begin{equation}\label{ep}
T=4\pi i \alpha \int \frac{d^4k}{(2\pi)^4}\;D_{im}(k)D_{jn}(k)
       \frac{\gamma_i({\hat l}-{\hat k}+m_e)\gamma_j}{k^2-2lk}\;M_{mn}.
\end{equation}
Here $D_{im}\,,\;\;D_{jn}$ are the photon propagators, we use the gauge 
$A_0=0$ for virtual photons; $l_{\mu}=(m_e,0,0,0)$ is the electron 
momentum. The nuclear-spin
independent Compton forward scattering amplitude, which is of
interest to as, can be written as
\begin{equation}\label{gp}
 M=\ba(\omega^2,{\bf k}^2){\bf E^*E}+\bb(\omega^2,{\bf k}^2){\bf B^*B}
 \;=M_{mn}e_me_n{}^*,
\end{equation}
where $\ba$ and $\bb$ are the nuclear electric and magnetic
polarizabilities, respectively. Both expression are exact (to the 
discussed order in the fine-structure constant $\alpha$).

We performed the integration in formula (\ref{ep}) with the logarithmic 
accuracy, i.e., neglecting in the result terms on the order of unity 
as compared to $\ln \bar E_p/m_e$. Here $\bar E_p$ is the typical
excitation energy for the proton, $m_e$ is the electron mass. It should
be emphasized that $\ln \bar E_p/m_e$ is not just a mere theoretical 
parameter,
it is truly large, about $6 - 7$, which makes the logarithmic
approximation quite meaningful quantitatively. This 
is the only theoretical approximation made in~\cite{ks}.

The resulting effective operator of the electron-proton interaction
(equal to $-T$) can be written in the coordinate representation as
\begin{equation}\label{d}
V=-\,\alpha m_e\, [\,5\ba(0)-\bb(0)\,]\;\ln\frac{\bar E}{m_e}
\;\delta({\bf r}).
\end{equation}
 
Now, we go over to the discussion of some statements made in~\cite{ro}.

\bigskip

1. The assertion of~\cite{ro}: ``For light electronic and muonic atoms
the energy shift due to virtual excitations can be written as an
integral over the forward virtual Compton amplitude'' is incorrect.  
In fact, this integral fixes (in the logarithmic approximation) the sum
of the polarizabilities $\ba(0)+\bb(0)$.  To calculate the 
correction, one needs also the data on 
the backward Compton scattering, which fix $\ba(0)-\bb(0)$.

\bigskip

2. The abstract and section {\bf 1.} of~\cite{ro} are 
written in such a way that an impression may arise that higher 
multipoles and transverse excitations have been neglected in our
paper~\cite{ks}.

In fact, higher multipoles are accounted for in formula (\ref{d}), as 
far as they contribute to the total photoabsorption cross section 
$\sigma_{\gamma}(\nu)$ and to the backward Compton scattering. 

As to transverse excitations, they are also accounted for in 
formula (\ref{d}). Indeed, in the combination $5\ba(0)-\bb(0)$, the 
contribution $4\ba(0)$ 
is due to the Coulomb interaction, but $\ba(0)-\bb(0)$ is due to the
exchange by transverse quanta, i.e., to the magnetic interaction of
convection currents~\cite{khr}.

\bigskip

3. The shift of the mean excitation energy $\bar E_p$ to 410 MeV, 
as advocated in~\cite{ro}, from our estimate 300 MeV, changes the 
result by 5\% only, which is well within our estimate 15\% for the
accuracy of the used logarithmic approximation. Obviously, such an 
improvement is not worth discussion.

\bigskip

4. We do not think that the mentioned estimate 15\% for our theoretical
accuracy can be improved by using ``the simple harmonic oscillator 
quark 
model'' of the nucleon, even if the quark mass $M_{quark}$ in it is 
taken with four digits (see the caption to Table 1 in~\cite{ro}).

\bigskip

5. The only, marginal, improvement over our results made in~\cite{ro}, 
is related to the use therein of more precise experimental data on the 
total photoabsorption cross section $\sigma_{\gamma}(\nu)$~\cite{ma}.
The value for the sum of the polarizabilities obtained from these data
in~\cite{ba} is
\beq\label{bab}
\ba_p(0)+\bb_p(0)=(13.69 \pm 0.14)\times 10^{-4}\; \mbox{fm}^3. 
\eeq
(Somewhat different number, without error bars, is presented 
in~\cite{ro}:
$\ba_p(0)+\bb_p(0)=13.75 \times 10^{-4}\; \mbox{fm}^3$.)
We used in~\cite{ks} an older value for $\sigma_{\gamma}(\nu)$, which 
resulted in 
\[ \ba_p(0)+\bb_p(0)=(14.2 \pm 0.5)\times 10^{-4}\; \mbox{fm}^3. \]
However, the uncertainty of our result originating from 
experimental error bars is
strongly dominated by the errors in the experimental data on the 
backward Compton scattering which give
\[ \ba_p(0)-\bb_p(0)=(10.0 \pm 1.8)\times 10^{-4}\; \mbox{fm}^3. \]
Thus, using (\ref{bab}) practically 
does not change the corresponding error, 7 Hz, of our final 
result. As to the central value of our result for the 
discussed correction to the hydrogen 1S state, it changes due 
to (\ref{bab}) 
slightly, from  -71 Hz to  -70 Hz, well within the mentioned
uncertainty of $\pm 7$ Hz.

\vspace*{1.0 cm}

We acknowledge the support by the Russian Foundation for Basic
Research through grant No. 98-02-17797, by the Ministry of Education
through grant No. 3N-224-98, by the Federal Program 
Integration -- 1998 through Project No. 274.


\begin{thebibliography}\\
\bibitem{ks}  I.B. Khriplovich and R.A. Sen'kov, nucl-th/9704043.
\bibitem{ks1} I.B. Khriplovich and R.A. Sen'kov,
              Phys.Lett. A {\bf 249}, 474 (1998); physics/9809022.
\bibitem{ro}  R. Rosenfelder, hep-ph/9903352.
\bibitem{khr} A.I. Milstein, S.S. Petrosyan, and I.B. Khriplovich,
              Zh.Eksp.Teor.Fiz. {\bf 109}, 1146, 1996 
              [Sov.Phys. JETP {\bf 82}, 616 (1996)].
\bibitem{ma}  M. MacCormic {\it et al.}, Phys. Rev.
              C {\bf 53}, 4127 (1996).               
\bibitem{ba}  D. Babusci, G. Giordano, and G. Matone, Phys. Rev.
              C {\bf 57}, 291 (1998).
                              
\end{thebibliography}
\end{document}